\documentclass{SciPost}
\usepackage{bbm}

\hypersetup{
    colorlinks,
    linkcolor={red!50!black},
    citecolor={blue!50!black},
    urlcolor={blue!80!black}
}

\usepackage[bitstream-charter]{mathdesign}
\DeclareSymbolFont{usualmathcal}{OMS}{cmsy}{m}{n}
\DeclareSymbolFontAlphabet{\mathcal}{usualmathcal}

\fancypagestyle{SPstyle}{
\fancyhf{}
\lhead{\colorbox{scipostblue}{\bf \color{white} ~SciPost Physics }}
\rhead{{\bf \color{scipostdeepblue} ~Submission }}

\fancyfoot[C]{\textbf{\thepage}}
}

\newcommand{\beq}{\begin{equation}}
\newcommand{\eeq}{\end{equation}}
\newcommand{\bea}{\begin{eqnarray}}
\newcommand{\eea}{\end{eqnarray}}
\newcommand{\nn}{\nonumber\\}
\newcommand{\eqlab}[1]{\label{eq:#1}}  
\newcommand{\eq}[1]{Eq.~(\ref{eq:#1})}
\newcommand{\Eq}[1]{Equation~(\ref{eq:#1})}
\newcommand{\eqs}[2]{Eqs.~(\ref{eq:#1}) and (\ref{eq:#2})}
\newcommand{\Eqs}[2]{Equations~(\ref{eq:#1}) and (\ref{eq:#2})}

\newcommand{\eqr}[2]{Eqs.~(\ref{eq:#1}-\ref{eq:#2})}

\newcommand{\figlab}[1]{\label{fig:#1}}  
\newcommand{\fref}[1]{Fig.~\ref{fig:#1}}
\newcommand{\Fref}[1]{Figure~\ref{fig:#1}}

\newcommand{\seclab}[1]{\label{sec:#1}} 
\newcommand{\sref}[1]{Sec.~\ref{sec:#1}}
\newcommand{\srefs}[2]{Secs.~\ref{sec:#1} and \ref{sec:#2}} 

\newcommand{\aref}[1]{Appendix~\ref{sec:#1}}

\newcommand{\ket}[1]{\vert#1\rangle}
\newcommand{\bra}[1]{\langle#1\vert}
\newcommand{\ip}[2]{\langle#1\vert#2\rangle}
\newcommand{\me}[3]{\langle#1\vert#2\vert#3\rangle}
\newcommand{\ev}[1]{\langle#1\rangle}
\newcommand{\w}{\omega}
\newcommand{\wpl}{\w_{\rm p}}

\newcommand{\mel}{m_{\rm e}}
\newcommand{\nel}{n_{\rm e}}

\newcommand{\Eg}{E_{\rm G}}
\newcommand{\Ep}{E_{\rm P}}
\newcommand{\El}{E_{\rm L}}
\newcommand{\Ry}{{\rm Ry}}

\renewcommand{\k}{{\bf k}}
\renewcommand{\r}{{\bf r}}

\newcommand{\q}{{\bf q}}
\newcommand{\R}{{\bf R}}
\newcommand{\bnk}{_{n\k}}
\newcommand{\bmk}{_{m\k}}
\newcommand{\0}{\mathbf{0}}
\renewcommand{\Re}{\mathrm{Re}}
\renewcommand{\Im}{\mathrm{Im}}
\begin{document}

\pagestyle{SPstyle}

\begin{center}{\Large \textbf{\color{scipostdeepblue}{
Optical bounds on many-electron localization\\
}}}\end{center}

\begin{center}\textbf{
Ivo Souza\textsuperscript{1,2},
Richard M.  Martin\textsuperscript{3,4} and
Massimiliano Stengel\textsuperscript{5,6}
}\end{center}

\begin{center}
{\bf 1} Centro de F{\'i}sica de Materiales, Universidad del Pa{\'i}s
Vasco, 20018 San Sebasti{\'a}n, Spain
\\
{\bf 2} Ikerbasque Foundation, 48013 Bilbao, Spain
\\
{\bf 3} Department of Physics, University of Illinois at
Urbana-Champaign,Urbana, Illinois 61801, USA
\\
{\bf 4} Department of Applied Physics, Stanford University, Stanford,
California 94305, USA
\\
{\bf 5} Institut de Ci\`encia de Materials de Barcelona (ICMAB-CSIC),
Campus UAB, 08193 Bellaterra, Spain
\\
{\bf 6} ICREA-Instituci\'o Catalana de Recerca i Estudis Avan\c{c}ats,
08010 Barcelona, Spain
\end{center}

\section*{\color{scipostdeepblue}{Abstract}} {\boldmath\textbf{We
    establish rigorous inequalities between different electronic
    properties linked to optical sum rules, and organize them into
    weak and strong bounds on three characteristic properties of
    insulators: electron localization length $\ell$ (the quantum
    fluctuations in polarization), electric susceptibility $\chi$, and
    optical gap $\Eg$.  All-electron and valence-only versions of the
    bounds are given, and the latter are found to be more informative.
    The bounds on~$\ell$ are particularly interesting, as they provide
    reasonably tight estimates for an ellusive ground-state property
    --~the average localization length of valence electrons~-- from
    tabulated experimental data: electron density, high-frequency
    dielectric constant, and optical gap.  The localization lengths
    estimated in this way for several materials follow simple chemical
    trends, especially for the alkali halides.  We also illustrate our
    findings via analytically solvable harmonic oscillator models,
    which reveal an intriguing connection to the physics of
    long-ranged van der Waals forces.  }}

\vspace{\baselineskip}

\noindent\textcolor{white!90!black}{%
\fbox{\parbox{0.975\linewidth}{%
\textcolor{white!40!black}{\begin{tabular}{lr}%
  \begin{minipage}{0.6\textwidth}%
    {\small Copyright attribution to authors. \newline
    This work is a submission to SciPost Physics. \newline
    License information to appear upon publication. \newline
    Publication information to appear upon publication.}
  \end{minipage} & \begin{minipage}{0.4\textwidth}
    {\small Received Date \newline Accepted Date \newline Published Date}%
  \end{minipage}
\end{tabular}}
}}
}


\vspace{10pt}
\noindent\rule{\textwidth}{1pt}
\tableofcontents
\noindent\rule{\textwidth}{1pt}
\vspace{10pt}


\section{Introduction}
\seclab{intro}

The low-frequency electronic conductivity,
\beq
\sigma_{aa}(\w) = \Re\, \sigma_{aa}(\w) + i\Im\,\sigma_{aa}(\w)\,,
\eeq
displays sharply different behaviors in metals and in insulators. To
characterize those behaviors one may define
\begin{subequations}
\begin{align}
D_{aa}&=\pi\lim_{\w\rightarrow 0}\, \w\,\Im\,\sigma_{aa}(\w)\,,\\
\epsilon_0\chi_{aa}&=-\lim_{\w\rightarrow 0}\, \w^{-1}\,\Im\,\sigma_{aa}(\w)\,,
\end{align}
\eqlab{drude-chi}%
\end{subequations}
where $\epsilon_0$ is the vacuum permittivity.  The Drude weight
$D_{aa}$ is finite in metals and vanishes in insulators, whereas the
clamped-ion electric susceptibility $\chi_{aa}$ is finite in
insulators and diverges in metals.  The $1/\w$ divergence of
$\Im\,\sigma_{aa}(\w)$ in perfect conductors is due to the
acceleration of free electrons under an applied electric field, and
its linear decrease with $\w$ in insulators reflects the polarization
of bound electrons in reaction to the field.

In 1964, Kohn proposed electron localization as the essential property
of the insulating state, and showed that it leads directly to its
distinctive electrical behavior~\cite{kohn-pr64}. He argued that the
ground-state wave function $\Psi(\r_1,\ldots,\r_N)$ of an insulator in
a periodic supercell breaks up into a sum of functions,
$\Psi=\sum_M\,\Psi_M$, which are localized in disconnected regions of
configuration space and have essentially vanishing overlap.  Kohn went
on to show that the disconectedness of $\Psi$ allows for the
definition of an effective center-of-mass operator ${\bf X}/N$, even
though the bare center-of-mass operator operator
$(1/N)\sum_{i=1}^N\,\r_i$ is ill-defined under periodic boundary
conditions.  The operator ${\bf X}$ is based on sawtooth functions,
whose discontinuities are placed in regions of configuration space
where $\Psi$ becomes exponentially small~\cite{kohn68}.

The importance of ${\bf X}$ can be seen from the fact that its
ground-state expectation value yields the electronic contribution to
the macroscopic electric polarization (${\bf P}$),
\beq
{\bf P}_{\rm e}=-|e|\ev{{\bf X}}/V\,,
\eqlab{Pel-def}
\eeq
where $V$ is the supercell volume.  Thanks to the development of the
modern theory of polarization, ${\bf P}$ is now understood as a
fundamental bulk property of crystalline insulators, independent of
surface termination modulo a discrete quantum of indeterminacy.  In
particular, within a single-particle band picture, \eq{Pel-def}
reduces to a sum over the Wannier centers (Kohn's disconnected wave
function pieces $\Psi_M$ can be viewed as ``many-body Wannier
functions''), or can be equivalently written as a Berry phase in
momentum space~\cite{king-smith-prb93,vanderbilt-book18}.  Crucially,
this theory asserts that bulk polarization is a property of the wave
function and not of the charge density, in line with Kohn's view on
electron localization.

In addition, Kohn's center-of-mass operator allows for the definition
of an electron localization tensor~\cite{souza-prb00}
\beq
\ell^2_{ab}=\frac{1}{N}
\left[
\langle X_a X_b \rangle -
\langle X_a \rangle \langle X_b \rangle
\right]\,.
\eqlab{loc-def}
\eeq
The diagonal entries of this symmetric tensor carry the interpretation
of a localization length squared, averaged over the total number of
electrons, along the corresponding direction. In high-symmetry
crystals, the localization tensor becomes isotropic:
\beq
\ell^2_{ab} = \delta_{ab} \ell^2\,.
\eqlab{loc-length}
\eeq
As in the case of ${\bf P}$, the localization tensor enjoys an elegant
formulation in the framework of band theory, where it can be written
as a quantum metric tensor~\cite{provost80} of the valence Bloch
manifold~\cite{resta-prl99,souza-prb00,souza-prb08}, whose Cartesian
trace is related to the Wannier spread~\cite{marzari-prb97}: see
\aref{band-insulators}.  First-principles studies of the localization
tensor have been carried out for tetrahedrally-coordinated
semiconductors~\cite{sgiarovello-prb01} and
oxides~\cite{veithen-prb02}.

Kohn did not directly relate the degree of wave function localization
to any physical observable.  An important step in that direction was
taken shortly before the modern theory of polarization was
developed. In Ref.~\cite{kudinov91}, Kudinov proposed to quantify
electron localization in insulators via the quantum fluctuations in
the ground-state polarization [\eq{loc-def}], connecting them to the
optical absorption spectrum by means of a fluctuation-dissipation
relation.  For a bulk crystal, such relation at zero temperature takes
the form~\cite{souza-prb00}
\beq
\ell^2_{ab}=
\frac{\hbar}{\pi e^2\nel}\int_0^\infty d\w\,
\w^{-1}\,\Re\,\sigma^{\rm S}_{ab}(\w)\,,
\eqlab{fluct-diss}
\eeq
where $\nel=N/V$ is the electron density, the superscript S denotes
the symmetric part of the conductivity tensor, and the integral spans
the positive-frequency optical absorption spectrum.\footnote{The
  conductivity tensor can be decomposed in three different ways: real
  and imaginary parts, $\Re\,\sigma$ and $\Im\,\sigma$; symmetric and
  antisymmetric parts, $\sigma^{\rm S}$ and $\sigma^{\rm A}$;
  Hermitian and anti-Hermitian parts, $\sigma^{\rm H}$ and
  $\sigma^{\rm AH}$. The Hermitian part of $\sigma$ (and hence
  $\Re\,\sigma^{\rm S}$) is dissipative, while the anti-Hermitian part
  is reactive.}  The trace of the localization tensor diverges in
conductors by virtue of their nonzero DC Ohmic conductivity, while in
insulators it remains finite.

The fluctuation-dissipation relation written above assumes a vanishing
macroscopic electric field ${\bf E}$, as appropriate for transverse
long-wave excitations.  The needed generalization to accomodate more
general electrical boundary conditions was given by
Resta~\cite{resta-prl06}. In particular, for longitudinal excitations
where ${\bf E}=-{\bf P}/\epsilon_0$ (${\bf D}=\0$), the
fluctuation-dissipation relation becomes a sum rule for the
energy-loss spectrum,
\beq
\tilde\ell^2(\hat\q)=
\frac{\hbar\epsilon_0}{\pi e^2\nel}\int_0^\infty d\w\,
\Im
\left[
-\epsilon^{-1}(\hat{\q},\w)
\right]\,,
\eqlab{fluct-diss-longitudinal}
\eeq
where the integrand is the (generally direction-dependent)
${\bf q}\rightarrow \0$ limit of the longitudinal inverse dielectric
function. The quantum fluctuations encoded in the localization tensor
depend on the electrical boundary conditions, and it is only under
${\bf E}=\0$ as assumed in \eq{fluct-diss} that its trace
discriminates between insulators and metals~\cite{resta-prl06}.  In
the following, we will deal mostly with the transverse localization
tensor; when referring to longitudinal quantities, we will denote them
with a tilde as done above.

\begin{table*}
\centering
\begin{tabular}{cccc}
\hline\hline
Length relations & References & Comments & Energy relations \\
\hline $\ell \leq \ell_{++}$ & \cite{souza-prb00} &
$\ell_{++}^2\propto 1/\Eg$  & $\El\ \geq \Eg$\\
& & Weak upper bound & \\
$\ell_- \leq \ell$ & \cite{aebischer-prl01,martin-book04} &
$\ell_-^2\propto \chi\Eg/\nel$ & $\Ep^2/\Eg \geq \El$ \\
& & Lower bound  & \\
& & Sum-rule derivation in~\cite{martin-book04} & \\
$\ell \leq \ell_+$ & \cite{verma2025,onishi2024quantum}, &
$\ell_+^2\propto\sqrt{\chi/\nel}$ & $\El \geq \Ep$\\
& this work & Strong upper bound & \\
$\ell_+ \leq \ell_{++}$ & This work &
Equivalent to $\ell_- \leq \ell_+,\ell_{++}$ & $\Ep\geq\Eg$\\
$\ell_- \leq \ell \leq \ell_+ \leq \ell_{++}$ & This work
& Chained inequalities &
$\Ep^2/\Eg \geq \El \geq \Ep \geq \Eg$
\\
\hline\hline
\end{tabular}
\caption{Overview of the sum-rule inequalities on $\ell$ discussed in
  the present work. Those inequalities relate the electron
  localization length $\ell$ defined by \eqs{loc-def}{loc-length} to
  the optical gap $\Eg$, the clamped-ion electric susceptibility
  $\chi$, and the electron density $\nel$. The last column contains
  equivalent energy relations involving the localization gap $\El$ and
  the Penn gap $\Ep$, which will be defined shortly [see \eq{El-Ep}].}
\label{tab:ineqs}
\end{table*}

Although \eq{fluct-diss} provides a way of extracting the transverse
localization length $\ell$ from the optical absorption spectrum, we
are not aware of any experimental work in that direction.  As
discussed in Ref.~\cite{martin-book04}, an alternative is to estimate
$\ell$ via rigorous upper and lower bounds involving readily-available
experimental data: electron density $\nel$, clamped-ion electric
susceptibility $\chi$, and minimum optical gap $\Eg$ (see
Table~\ref{tab:ineqs}).  This strategy was used recently to estimate
$2\pi\nel\ell^2$ (the quantum metric of the filled bands) for a number
of materials~\cite{komissarov-natcomms24,onishi2024quantum}.

In this work, we employ a sum-rule approach to establish weak and
strong bounds on $\ell$, $\chi$, and $\Eg$.  We give two formulations
of the bounds --~all electron and valence-only~-- and argue that the
valence-only formulation, even if approximate, is more
informative. This is confirmed by an explicit evaluation of the bounds
on $\ell$ for a series of materials; the strong bound is found to be
much tighter than the weak one, and the valence-only formulation
reveals simple chemical trends.  To illustrate the impact of
long-ranged electrostatics on the polarization
fluctuations~\cite{resta-prl06}, we apply our formalism to
analytically solvable systems of harmonic oscillators. This exercise
reveals an intriguing connection to the physics of van der Waals
(dispersion) forces, and clarifies the central role of
electron-electron correlation in the determination of the optical
bounds.

The manuscript is organized as follows. In \sref{sum-rules} the
inverse moments of the optical absorption spectrum are introduced, the
sum rules for the three leading moments are stated, and average
optical gaps are defined. In \sref{bounds}, sum-rule inequalities are
established for the inverse moments and for the average gaps; the
latter are then organized into chained inequalities, from which
various bounds on $\ell$, $\chi$, and $\Eg$ are deduced. In
\sref{analytic} those bounds are examined for several exactly-solvable
models, including harmonic oscillator models coupled by dispersion
interactions.  In \sref{estimation} the localization length $\ell$ is
estimated for several materials using the all-electron and
valence-only varieties of the bounds, and the observed trends are
discussed. We conclude in \sref{conclusions} with a summary, and
provide some accessory results in three appendices.

\section{Sum rules and average gaps}
\seclab{sum-rules}

For light with linear polarization along direction $\hat{\bf n}$, we
define the inverse moments of the positive-frequency optical
absorption spectrum at zero temperature as
\beq
I_p(\hat{\bf n})=\frac{2}{\pi}\int_0^\infty d\w\,
\w^{-p}\,\Re\,\sigma^{\rm S}_{ab}(\w)\,
\hat n_a \hat n_b\,,
\eqlab{Ip-def}
\eeq
where $p \geq 0$, a summation over repeated Cartesian indices is
implied, and the $2/\pi$ factor was included for convenience in
writing the sum rules below.  For simplicity we will assume cubic
symmetry or higher so that
$\sigma^{\rm S}_{ab}=\delta_{ab}\sigma^{\rm S}$, rendering $I_p$
independent of~$\hat{\bf n}$,
\beq
I_p=\frac{2}{\pi}\int_0^\infty d\w\,
\w^{-p}\,\Re\,\sigma^{\rm S}(\w)\,.
\eqlab{Ip-cubic}
\eeq

The inverse spectral moments with $p=0,1,2$ satisfy
\begin{subequations}
\begin{align}
I_0&=\frac{e^2\nel}{\mel}\equiv\epsilon_0\wpl^2\,,
\eqlab{I0-SR}\\
I_1&=\frac{2e^2}{\hbar}\nel\ell^2\,,
\eqlab{I1-SR}\\
I_2&=\epsilon_0\chi
\equiv\epsilon_0\left(\epsilon-1\right)\,,
\eqlab{I2-SR}
\end{align}
\eqlab{sum-rules}%
\end{subequations}
where we have introduced the static electronic permittivity $\epsilon$
(often denoted as $\epsilon_\infty$), and the plasma frequency
$\wpl$.\footnote{In general, the frequency $\wpl$ defined by
  \eq{I0-SR} does not correspond to a physical resonance of the
  medium. The physical meaning of the parameter $\wpl$ is provided by
  the free-electron-like behavior of the dielectric function at
  frequencies far above the deepest core-level
  resonance~\cite{jackson-book99,landau-book84}:
  $\epsilon(\w)/\epsilon_0 \simeq 1-\wpl^2/\w^2$. In a real plasma,
  electrons are free and the range of validity of this formula is very
  broad, including $\w<\wpl$~\cite{jackson-book99}. In that case,
  $\wpl$ does correspond to a physical resonance of the medium.}  The
above identities are respectively the oscillator-strength sum rule,
the fluctuation-dissipation relation of \eq{fluct-diss} [with $\ell^2$
given by \eq{loc-length}], and the electric-susceptibility sum rule.
All three sum rules converge for insulators, while in metals $I_1$ and
$I_2$ diverge as a result of the nonzero DC conductivity.
\Eqs{I0-SR}{I2-SR} follow from the Kramers-Kr\"onig relations, which
in the case of~\eqref{eq:I0-SR} must be combined with the observation
that at sufficiently high frequencies the medium responds to an
electromagnetic disturbance like a free-electron
gas~\cite{landau-book84,wooten-book72}.  The corresponding sum rules
for atomic systems are well
known~\cite{fano-rmp68,bethe-book86,traini-ejp96}. In solid-state
physics, atomic-like sum rules have been used to characterize F
centers in alkali halide crystals~\cite{brauwers-prb75}; in
particular, from the ratio between the $I_1$ and $I_0$ moments of the
F-center absorption band one can deduce, under the effective-mass
approximation, its mean radius in the ground state.

To proceed, we find it useful to define a ``localization gap'' $\El$
and a ``Penn gap'' $\Ep$ as~\cite{martin-book04}
\beq
\El^{-1}=\frac{\hbar^{-1}I_1}{I_0}\,,
\qquad
\Ep^{-2}=\frac{\hbar^{-2}I_2}{I_0}\,.
\eqlab{El-Ep}
\eeq
These average inverse excitation energies weighted by the transition
strength~\cite{traini-ejp96} will be denoted as (inverse) ``average
gaps.''  Using \eq{sum-rules} and writing $\hbar^2/2\mel$ as
$a_0^2\,\Ry$ ($a_0$ is the Bohr radius and Ry is the Rydberg unit of
energy), we obtain
\beq
\El=\frac{\hbar^2}{2\mel\ell^2}
\Leftrightarrow
\left( \frac{\ell}{a_0} \right)^2 = \frac{\Ry}{\El}
\eqlab{loc}
\eeq
and
\beq
\chi=\left( \frac{\hbar\wpl}{\Ep}\right)^2\,,
\eqlab{chi}
\eeq
the latter being the standard definition of the Penn gap in
semiconductor physics~\cite{penn-pr62,yu-book10}.

As shown below and already indicated in Table~\ref{tab:ineqs}, the
inequalities of interest can be expressed concisely as relations among
three characteristic energy scales of the band structure: optical gap
$\Eg$ (the energy threshold for optical absorption), Penn gap, and
localization gap.

\section{Sum-rule inequalities}
\seclab{bounds}

If the unperturbed system is in thermodynamic equilibrium, we
have~\cite{landau-book84}
\beq
\sigma^{\rm S}(\w) \geq 0 \quad \text{for $\w>0$}\,.
\eqlab{equil}
\eeq
From this condition, one can readily obtain two types of inequalities
involving different spectral moments~\cite{traini-ejp96}. The first
type are of the form
\beq
I_{p+q}\leq\frac{\hbar}{\Eg}I_{p+q-1}\leq\ldots\leq
\left( \frac{\hbar}{\Eg} \right)^q I_p\,,
\eqlab{Eg-ineqs}
\eeq
where $q>0$; the second,
\beq
I_p^2\leq I_{p-1}I_{p+1}\,,
\eqlab{cauchy-ineq}
\eeq
follow from the Cauchy-Bunyakovsky-Schwarz inequality
\beq
\left(
\int_0^\infty d\w\,f(\w)g(\w)
\right)^2
\leq
\left(
\int_0^\infty d\w\, f(\w)^2
\right)
\left(
\int_0^\infty d\w\, g(\w)^2
\right)
\eeq
by setting $f(\w)=\w^{-(p-1)/2}\sqrt{\Re\,\sigma^{\rm S}(\w)}$ and
$g(\w)=\w^{-(p+1)/2}\sqrt{\Re\,\sigma^{\rm S}(\w)}$.  Both types of
inequalities saturate in the limit of a narrow absorption spectrum
concentrated at $\Eg$~\cite{traini-ejp96}.

The average gaps introduced in \eq{El-Ep} satisfy
\beq
\El\geq\Ep\geq\Eg\,,\qquad \Ep^2\geq\Eg\El\,,
\eqlab{El-Ep-ineqs}
\eeq
with the relation $\El\geq\Ep$ coming from \eq{cauchy-ineq} and the
others from \eq{Eg-ineqs}; as expected, the average gaps $\El$ and
$\Ep$ cannot be smaller than the minimum gap $\Eg$.
\Eq{El-Ep-ineqs} allows to bracket $\El$ as
$\Ep^2/\Eg\geq\El\geq\Ep\geq\Eg$ and $\Ep$ as
$\El^2\geq \Ep^2\geq \Eg\El\geq \Eg^2$; combined with \eqs{loc}{chi},
these chained inequalities yield
\begin{subequations}
\begin{align}
\frac{\epsilon_0\chi\Eg}{2e^2\nel}
&\leq
\ell^2
\leq
\frac{\hbar}{2|e|}
\sqrt{\frac{\epsilon_0\chi}{\mel\nel}}
\leq
\frac{\hbar^2}{2\mel\Eg}\qquad
\left(
\ell^2_- \leq \ell^2 \leq \ell^2_+ \leq \ell^2_{++}
\right)\,,
\eqlab{bounds-loc}\\
\frac{4e^2\mel\nel\ell^4}{\epsilon_0\hbar^2}
&\leq
\chi
\leq
\frac{2e^2\nel\ell^2}{\epsilon_0\Eg}
\leq
\frac{\hbar^2 e^2\nel}{\epsilon_0\mel\Eg^2}\,.
\eqlab{bounds-chi}
\end{align}
\eqlab{bounds-loc-chi}%
\end{subequations}
We will refer to $\ell_-$ as the lower bound on $\ell$, and to
$\ell_+$ and $\ell_{++}$ as the strong and weak upper bounds,
respectively; the same terminology will be used for the bounds on
$\chi$. The weak upper bounds on $\ell$~\cite{souza-prb00} and on
$\chi$~\cite{onishi-prb24} reflect the intuitive notion that wide-gap
materials tend to have more localized and less polarizable electrons.

The bounds on $\ell$ are particularly interesting, as they only
involve parameters that are tabulated for many materials: electron
density, electric susceptibility, and optical gap.  Since $\ell$
itself is not commonly measured, those bounds provide a simple and
practical way of estimating its value.  Note that the weak upper bound
$\ell_{++}$ only depends on the inverse minimum gap; this is a
delicate quantity, especially for narrow-gap semiconductors, and it is
not representative of the entire spectrum (the nature of the electron
system can be very different for materials with the same minimum gap).
The localization length is instead a global property of the electron
system, and the value of $\Eg$ is not its most relevant descriptor;
for example, $\ell_{++}$ diverges in the same way for all materials as
$\Eg$ is tuned to zero. We therefore expect $\ell_{++}$ to give a
relatively poor estimate for $\ell$ in real systems. The strong upper
bound $\ell_+$ depends instead on $\chi$ and $\nel$ via the average
Penn gap, which is much more representative of the entire spectrum.
As for the lower bound $\ell_-$, it depends on both $\Eg$ and $\Ep$;
there is still some dependence on the minimum gap, but it is a smaller
effect than for $\ell_{++}$.

The relations in \eq{El-Ep-ineqs} can also be arranged as
$\Eg \leq \Ep^2/\El \leq \Ep \leq \El$ to place bounds on the optical
gap,
\beq
\Eg
\leq
\frac{2e^2\nel\ell^2}{\epsilon_0\chi}
\leq
\hbar|e|
\sqrt{\frac{\nel}{\mel\epsilon_0\chi}}
\leq
\frac{\hbar^2}{2\mel\ell^2}\,.
\eqlab{bounds-gap}
\eeq
It is significant that there are several upper bounds, but no lower
bound. This is consistent with the existence of electronic systems
without an energy gap that are strict
insulators~\cite{kohn-pr64,kohn68}.

Although our focus has been on transverse long-wave modes, similar
results hold for longitudinal
modes~\cite{onishi-prb24,onishi2024quantum}.  The only changes to
\eqs{bounds-loc-chi}{bounds-gap} are (see \aref{longitudinal})
\beq
\Eg \rightarrow \tilde E_{\rm G}\,,
\qquad
\ell\rightarrow \tilde\ell\,,
\qquad
\chi\rightarrow 1-\epsilon^{-1}\,,
\eqlab{changes-LO}
\eeq
where $\tilde E_{\rm G}$ is the minimum energy for long-wave
longitudinal excitations (plasmon gap), and $\tilde\ell$ was
introduced in \eq{fluct-diss-longitudinal} (in high-symmetry crystals,
$\tilde\ell$ does not dependend on $\hat\q$). The lower and strong
upper bounds on $\tilde\ell^2$ are given in
Ref.~\cite{onishi2024quantum} (in terms of $2\pi\nel\tilde\ell^2$),
and the weak upper bound on $1-\epsilon^{-1}$ is given in
Ref.~\cite{onishi-prb24}.

In closing, we comment on the applicability of the above relations to
Chern insultors (CIs).  The general character of the sum rules in
\eq{sum-rules} suggests that the inequalities deduced from them remain
valid for CIs. The subtlety is that CIs occupy a middle ground between
metals and ordinary insulators~\cite{vanderbilt-book18}, and the $I_1$
and $I_2$ sum rules diverge for metals. On the other hand, all three
sum rules involve the symmetric (time even) part of the optical
conductivity, whereas the distinction between ordinary and Chern
insulators rests with the antisymmetric (time odd) part; from this we
can conclude that the inequalities obtained above do apply to CIs,
even if such materials fall outside the scope of Kohn's theory of the
insulating state.  Indeed, while the total Wannier spread diverges in
a CI, its gauge invariant part proportional to $\ell^2$ remains
finite~\cite{thonhauser-prb06}, consistent with the weak upper bound
on $\ell^2$.  Likewise, the weak upper bound on~$\chi$ implies that
the susceptibility remains finite in CIs, even though the concept of
spontaneous polarization requires special care~\cite{coh-prl09}.

\section{Analytically solvable models}
\seclab{analytic}

To build intuition on the bounds obtained above, we will now apply
them to several models that can be treated analytically.  For the
first few examples dealing with finite systems, we introduce a
polarizability per electron via the relation
${\bf d}=N\alpha{\bf E}_0$; here ${\bf d}$ is the dipole moment
induced on the $N$-electron system by the applied electric field
${\bf E}_0$.  To use the bulk relations~\eqref{eq:bounds-loc-chi}
and~\eqref{eq:bounds-gap}, we place the system in a periodic
supercell.  In the limit where the supercell dimensions far exceed
those of the system, the applied field ${\bf E}_0$ generates a
macroscopic field ${\bf E}={\bf E}_0$ in the effective medium; from
${\bf P}=\epsilon_0\chi{\bf E}={\bf d}/V$ we get
$\chi=\nel\alpha/\epsilon_0+{\cal O}(V^{-1})$, where the additional
terms (originating from the Clausius-Mossotti relation, see
\sref{vdw-xtal}) vanish in the assumed limit of large $V$.  Plugging
this expression for $\chi$ into \eqs{bounds-loc-chi}{bounds-gap} gives
\begin{subequations}
\begin{align}
\frac{\alpha\Eg}{2e^2}
&\leq
\ell^2
\leq
\frac{\hbar}{2|e|}\sqrt{\frac{\alpha}{\mel}}
\leq
\frac{\hbar^2}{2\mel\Eg}\,,\\
\frac{4 e^2 \mel \ell^4}{\hbar^2}
&\leq
\alpha
\leq
\frac{2e^2\ell^2}{\Eg}
\leq
\frac{\hbar^2e^2}{\mel \Eg^2}\,,
\eqlab{bounds-chi-finite}\\
\Eg
&\leq
\frac{2e^2\ell^2}{\alpha}
\leq
\frac{\hbar|e|}{\sqrt{\mel\alpha}}
\leq
\frac{\hbar^2}{2\mel\ell^2}\,.
\end{align}
\eqlab{bounds-finite}
\end{subequations}

At this point we make contact with known results for atoms and
molecules.  The strong upper bound on $\alpha$, with $\Eg$ replaced by
a mean excitation energy $\Delta E$ and $3e^2\ell^2$ expressed as the
dipole fluctuation $\langle d^2 \rangle - \langle d \rangle^2$,
becomes
\beq
\alpha\approx \frac{2}{3}\frac{\langle d^2 \rangle - \langle d \rangle^2}
{\Delta E}\,.
\eeq
This estimate for the polarizability is discussed in
Ref.~\cite{atkins-book05}, along with its relation to the
fluctuation-dissipation relation.  That textbook also gives an
estimate for $\alpha$ in terms of the weak upper bound in
\eq{bounds-chi-finite}, invoking the oscillator-strength sum rule.

\subsection{Hydrogen atom}

Introducing the polarizability volume
$\alpha'=\alpha/4\pi\epsilon_0$~\cite{atkins-book05},
\eq{bounds-finite} becomes
\begin{subequations}
\begin{align}
\frac{1}{4}\frac{\alpha'}{a_0^3}\frac{\Eg}{\Ry}
&\leq
\frac{\ell^2}{a_0^2}
\leq
\frac{1}{2}\sqrt{\frac{\alpha'}{a_0^3}}
\leq
\frac{\Ry}{\Eg}\,,\\
4\frac{\ell^4}{a_0^4}
&\leq
\frac{\alpha'}{a_0^3}
\leq
4\frac{\Ry}{\Eg}\frac{\ell^2}{a_0^2}
\leq
4\frac{\Ry^2}{\Eg^2}\,,\\
\frac{\Eg}{\Ry}
&\leq
4\frac{a_0\ell^2}{\alpha'}
\leq
2\sqrt{\frac{a_0^3}{\alpha'}}
\leq
\frac{a_0^2}{\ell^2}\,,
\end{align}
\eqlab{bounds-atomic}%
\end{subequations}
where every fraction is dimensionless.  For the nonrelativistic
hydrogen atom we have~\cite{sakurai-book94,landau-book77}
\beq
\Eg=0.75\Ry\,,
\qquad
\alpha'=4.5a_0^3\,,
\qquad
\ell^2=a_0^2\,,
\eeq
which plugged into \eq{bounds-atomic} gives
\begin{subequations}
\begin{align}
\frac{27}{32}
&\leq
\frac{\ell^2}{a_0^2}=1
\leq
\sqrt{\frac{9}{8}}
\leq
\frac{4}{3}\,,\\
4
&\leq
\frac{\alpha'}{a_0^3}=4.5
\leq
\frac{16}{3}
\leq
\left( \frac{8}{3} \right)^2\,,
\eqlab{bounds-chi-atomic}
\\
\frac{\Eg}{\Ry}=0.75
&\leq
\frac{8}{9}
\leq
\frac{2}{\sqrt{4.5}}
\leq
1\,.
\end{align}
\end{subequations}
The lower and strong upper bounds on $\alpha'$ are given in
Ref.~\cite{traini-ejp96}, and the latter is also discussed in
Ref.~\cite{sakurai-book94} and in other textbooks. Interestingly, both
bounds can be improved by means of correction terms involving positive
moments of the absorption spectrum~\cite{traini-ejp96}.

Taking the average of the lower and strong upper bounds on $\ell^2$
and on $\alpha'$ produces the reasonably accurate estimates
$\ell^2\approx 0.952 a_0^2$ and $\alpha'\approx 4.(6) a_0^3$.  The
estimates $\ell^2\approx 1.089 a_0^2$ and $\alpha'\approx 5.(5) a_0^3$
obtained by taking the average of the lower and weak upper bounds are
much less accurate, especially for $\alpha'$.  We also note that the
strong upper bound $\ell_+$ is closer to $\ell$ than the lower bound
$\ell_-$.  This supports the notion that $\ell_+$, being based solely
on the average Penn gap, is more representative of the entire
absorption spectrum than $\ell_-$, which also depends on the minimum
gap. Further evidence that $\ell_+$ tends to track~$\ell$ more closely
than $\ell_-$ will be presented in \sref{estimation} for crystalline
materials.

\subsection{Isotropic harmonic oscillator}

For an electron trapped in an isotropic harmonic potential of
frequency $\w_0$ the parameters
are~\cite{sakurai-book94,atkins-book05}
\beq
\Eg=\hbar\w_0\,,\quad
\alpha=\frac{e^2}{\mel\w_0^2}\equiv\alpha_0\,,\quad
\ell^2=\frac{\hbar}{2\mel\w_0}\equiv\ell^2_0\,,
\eqlab{params-SHO}
\eeq
saturating all the inequalities in \eq{bounds-finite}. This can be
understood from the selection rules for the harmonic oscillator: as
the only allowed dipole transition from the ground state is to the
first excited state, the entire spectral weight is at $\Eg$, producing
the saturation.

\subsection{Van der Waals dimer model}
\seclab{vdw-dimer}

So far we have only discussed one-electron systems.  To analyze the
effect of electron correlations, we now consider a system of two
identical harmonic oscillators~1 and~2 separated by $\R$.  We think of
these oscillators as vibrating electrical dipoles in which the $+e$
charges (ions) are held in the position of equilibrium while the $-e$
charges (electrons) vibrate about these equilibrium positions, their
displacements being ${\bf r}_1$ and ${\bf r}_2$.  In the limit where
$r_1,\,r_2\ll R$, this provides a simple model for the van der Waals
interaction~\cite{atkins-book05,kittel-book04}.

The interaction term is
\beq
H_{12} =
\frac{e^2}{4\pi\epsilon_0}
\left(
\frac{1}{R} + \frac{1}{|\R+\r_1- \r_2|}
- \frac{1}{|\R+\r_1|}
-\frac{1}{|\R - \r_2|}
\right)
\,.
\eqlab{H12-full}
\eeq
In the approximation $r_1,\,r_2\ll R$ we expand \eq{H12-full} to
obtain in lowest order
\beq
H_{12} \simeq
\frac{e^2}{4\pi\epsilon_0}
r_{1a} r_{2b}
\left(\frac{\delta_{ab}}{R^3} - 3 \frac{R_a R_b}{R^5} \right)\,,
\eqlab{H12-expanded}
\eeq
which is in the form of a dipole-dipole interaction.  Orienting the
Cartesian frame such that $\R=R\hat{\bf x}$ leads to
\begin{align}
H_{12}&\simeq\frac{e^2}{4\pi\epsilon_0}
\left(
-\frac{2}{R^3}x_1x_2 + \frac{1}{R^3}y_1y_2+ \frac{1}{R^3}z_1z_2
\right)
\nn
&\equiv H^\parallel_{12} + H^\perp_{12}\,,
\eqlab{H12-expanded-b}
\end{align}
where $H^\parallel_{12}$ denotes the first term and $H^\perp_{12}$ the
other two.

For oscillations along $\R$ the only surviving term in
\eq{H12-expanded-b} is $H_{12}^\parallel$, and we recover the 1D model
of Ref.~\cite{kittel-book04}.  Denoting by $H_0$ the Hamiltonian of
the two uncoupled oscillators, $H_0+H^\parallel_{12}$ is diagonalized
by the transformation
\beq
x_\pm=\frac{1}{\sqrt{2}}(x_1\pm x_2)
\eeq
together with a similar transformation for the momenta, resulting in
two decoupled oscillators with frequencies
\beq
\w_\pm^\parallel
=\w_0\sqrt{1\mp\frac{2\alpha'_0}{R^3}}\,,
\eqlab{w-parallel}
\eeq
where $\alpha'_0$ is the polarizability volume of a single oscillator.

For unrestricted 3D oscillations, the interaction term is given by the
full \eq{H12-expanded-b}. Now instead of two modes we have six modes.
By following through the same derivation, we can split also the $y$
and $z$ modes into symmetric and antisymmetric combinations with
frequencies
\beq
\w_\pm^\perp=
\w_0\sqrt{1\pm\frac{\alpha'_0}{R^3}}
\,;
\eqlab{w-perp}
\eeq
thus, for transverse oscillations the symmetric modes have higher
frequency than the antisymmetric ones.

In the 3D model the parameters $\Eg$, $\alpha$, and $\ell^2$ are
anisotropic, carrying labels $\parallel$ or $\perp$.  To evaluate the
$\parallel$ components, note that the interaction with a field
${\bf E}_\parallel=E\hat{\bf x}$ is described by
$eE(x_1+x_2)=\sqrt{2}eEx_+$, and that $\ell_\parallel^2$ is defined
via~\eqref{eq:loc-def} in terms of $X\equiv x_1+x_2=\sqrt{2}x_+$. This
means that only the symmetric mode participates, and with a simple
calculation one finds that the three parameters are obtained by
replacing $\w_0$ with $\w^\parallel_+$ in
\eq{params-SHO},
\begin{subequations}
\begin{align}
E_{\rm G}^\parallel &= \hbar\w_+^\parallel\simeq
\hbar\w_0\left( 1 - \alpha'_0/R^3 \right)\,,\\
\alpha_\parallel &= \frac{e^2}{\mel\left(\w_+^\parallel\right)^2}\simeq
\alpha_0\left( 1 + 2\alpha'_0/R^3 \right)\,,\\
\ell^2_\parallel &= \frac{\hbar}{2\mel\w_+^\parallel}\simeq
\ell_0^2\left( 1 + \alpha'_0/R^3 \right)\,.
\end{align}
\eqlab{vdw-dimer}%
\end{subequations}
The $\perp$ components are obtained by sending
$\w_+^\parallel\rightarrow\w_+^\perp$ and
$\alpha'_0\rightarrow -\alpha'_0/2$ in these expressions.

In conclusion, the van der Waals interaction reduces the optical gap
and increases both the polarizability and the localization length in
the axial direction of the dimer, and the opposite happens in the
perpendicular directions.  As the antisymmetric modes are
dipole inactive, the entire spectral weight for light polarized along
$\R$ or perpendicularly to it is concentrated at a single frequency
$\hbar\w_+^\parallel$ or $\hbar\w_+^\perp$, respectively.  In both
cases the bounds in \eq{bounds-finite} remain saturated, just like for
a single oscillator.

We emphasize that the explicit treatment of electron correlations is
essential to obtain a qualitatively correct physical picture. For
example, it is easy to show that the fluctuation-dissipation sum rule
fails if the electron-electron interaction is treated at the
mean-field level, e.g., within Hartree-Fock theory. Within
Hartree-Fock, the dielectric susceptibility of the system of
interacting oscillators is described exactly; nonetheless, the
localization length is unaffected by the interaction and corresponds
to that of the isolated monomer. This implies that the correct
description of the macroscopic polarization fluctuations goes hand in
hand with the ability of the theory to capture dispersion interactions
between isolated bodies.

\subsection{Van der Waals crystal model}
\seclab{vdw-xtal}

As an extension of the dimer model, we now consider a periodic array
of oscillators coupled by dipole-dipole interactions.  The potential
energy reads
\beq
U = \frac{1}{2}\mel\w_0^2\sum_\R\,\left| \r^\R \right|^2
+
\frac{e^2}{4\pi\epsilon_0}
\sum_\R\sum_{\R'\neq \0} \frac{r^\R_{a} r^{\R+\R'}_{b}}{2} 
  \left(\frac{\delta_{ab}}{{R'}^3} - 3 \frac{R'_a R'_b}{{R'}^5} \right)\,,
\eqlab{U}
\eeq
where $\r^\R$ denotes the displacement of an electron away from its
equilibrium position $\R$, taken to be a point on a Bravais lattice.
A similar model was discussed in Ref.~\cite{onishi-prb24}; the
only difference is that the positive charges, instead of being point
charges placed at the lattice points, are smeared into a uniform
background.

\subsubsection{Dynamical matrix}

As in the dimer model, the electrons are assumed to be strongly
localized in the sense that the quantum fluctuations are small
compared to the separation between the ions.  The resulting potential
bears many similarities to the form that appears in the context of
lattice vibrations; we will therefore borrow the same terminology in
discussing the relevant contributions to the electronic Hamiltonian.
  
To determine the normal modes of the system we first evaluate the
force-constant matrix
\begin{align}
D_{a\R,b\R'} \equiv \frac{\partial^2 U}{\partial r_a^\R \partial r_b^{\R'}} =
D_{a\0,b\R'-\R}
\end{align}
to find
\beq
D_{a\0,b\R} = \mel\w_0^2\delta_{ab}\delta_{\R\0}
+\frac{e^2}{4\pi\epsilon_0} \left( 1-\delta_{\R\0} \right)
\left(
\frac{\delta_{ab}}{R^3} - 3\frac{R_a R_b}{R^5}
\right)\,,
\eeq
and then convert it into a dynamical matrix using
\beq
D_{ab}(\q) = \frac{1}{\mel}
\sum_\R\, D_{a\0,b\R}\, e^{-i\q\cdot\R}\,.
\eeq
The result is
\beq
D_{ab}(\q) = \w_0^2\delta_{ab} + C_{ab}(\q)\,,
\eeq
where
\beq
C_{ab}(\q) = \frac{e^2/\mel}{4\pi\epsilon_0}
\sum_{\R\not=\0}\, e^{-i\q\cdot\R}
\left(
\frac{\delta_{ab}}{R^3} - 3\frac{R_a R_b}{R^5}
\right)\,.
\eqlab{C-q}
\eeq

To carry out the above lattice sum it is convenient to work in
reciprocal space, where the interaction can be recast as a rapidly
converging Ewald summation,
\begin{equation}
C_{ab}({\bf q}) =
\frac{e^2/\mel}{4\pi\epsilon_0}
\left(
\frac{4\pi}{\Omega} \sum'_{\bf G} \frac{K_aK_b}{K^2} e^{-\frac{K^2 \sigma^2}{4}} - 
\delta_{ab} \frac{4}{3 \sqrt{\pi} \sigma^3}
\right)
\,, \qquad {\bf K}={\bf G+q}\,,
\eqlab{C-q-ewald}
\end{equation}
with $\Omega$ the volume of a primitive cell. The primed sum excludes
the divergent ${\bf G+q}=\0$ term, and the second term removes the
self-interaction of the dipole in the origin cell; the result is
independent of the Ewald parameter $\sigma$ provided that
$\sigma \ll R$ for all ${\bf R}\neq \0$.

By diagonalizing the $3\times 3$ matrix $C({\bf q})$ at every point in
the Brillouin zone, we have rewritten the problem as a set of
independent oscillators. In particular, we have three modes at each
${\bf q}$ with frequencies
\begin{equation}
\w^2_i({\bf q}) = \w^2_0 + \lambda_i({\bf q})\,,
\eqlab{omega-q}
\end{equation}
where $\lambda_i({\bf q})$ are the eigenvalues of $C({\bf q})$.

In \aref{vdw-zpe}, we calculate the zero-point energy of this model by
collecting the contributions from all normal modes across the
Brillouin zone.

\subsubsection{Long-wave limit}
\seclab{long-wave}

The $\q \rightarrow \0$ limit is particularly relevant to our
discussion, since it corresponds to the collective displacement of the
electronic center of mass.  For a cubic lattice we find two TO modes
and one LO mode where
\beq
\lambda_{\rm TO} =  -\frac{1}{3} \wpl^2\,, \qquad
\lambda_{\rm LO} =  \frac{2}{3} \wpl^2\,,
\eqlab{lambda-TO-LO}
\eeq
with $\wpl^2=e^2/\epsilon_0\mel\Omega$.  To obtain this result note
that the matrix $C(\q)$ is traceless so that
$2\lambda_{\rm TO}+\lambda_{\rm LO}=0$, and that
$\lambda_{\rm LO}=\lambda_{\rm TO}+\wpl^2$, where $\wpl^2$ is the
contribution from the ${\bf G}=\0$ term in \eq{C-q-ewald} when
$\q\rightarrow\0$.

The dielectric susceptibility and permittivity are readily given in
terms of the TO mode frequency,
\begin{equation}
\chi = \frac{\wpl^2}{\w_{\rm TO}^2}\,,
\qquad
\epsilon = 1 + \chi\,.
\end{equation}
Then, based on the above, we can quickly verify that the following
results hold,
\beq
\epsilon = \frac{\w_{\rm LO}^2}{\w_{\rm TO}^2}\,,
\qquad
\frac{\epsilon-1}{\epsilon+2} = \frac{\alpha_0}{3\epsilon_0\Omega}\,.
\eeq
The first result is the Lyddane-Sachs-Teller
relation~\cite{kittel-book04}, valid for a single-mode dielectric. The
second is the Clausius-Mossotti relation~\cite{kittel-book04}, linking
the macroscopic permittivity to the molecular polarizability
$\alpha_0$.

For the TO modes we have
\beq
\Eg=\hbar\w_{\rm TO}\,,
\qquad
\chi=\frac{\wpl^2}{\w^2_{\rm TO}}\,,
\qquad
\ell^2=\frac{\hbar}{2\mel\w_{\rm TO}}\,.
\eeq
When plugged into \eqs{loc}{chi} these parameters give $\El=\Ep=\Eg$,
saturating all the bounds in \eqs{bounds-loc-chi}{bounds-gap}.  The
parameters for the LO modes are
\beq
\tilde E_{\rm G}=\hbar\w_{\rm LO}\,,
\qquad
1-\epsilon^{-1}=\frac{\wpl^2}{\w^2_{\rm LO}}\,,
\qquad
\tilde \ell^2=\frac{\hbar}{2\mel\w_{\rm LO}}\,,
\eeq
and again the corresponding bounds, obtained by modifying
\eqs{bounds-loc-chi}{bounds-gap} according to \eq{changes-LO}, are
saturated.

\section{ Real materials}
\seclab{estimation}

Starting from experimental data, we have evaluated the bounds on
$\ell$ in \eq{bounds-loc} for a number of materials. To visualize the
results, it is helpful to bring that equation to the form
\beq
\frac{\Ry}{\Ep}\sqrt{\frac{\Eg}{\Ry}}
\leq
\frac{\ell}{a_0}
\leq
\sqrt{\frac{\Ry}{\Ep}}
\leq
\sqrt{\frac{\Ry}{\Eg}}
\,,
\eqlab{bounds-loc-fig}
\eeq
which suggests plotting the data as shown schematically in
\fref{schematic}.  Given a data point (large blue dot), the range
$[\ell_-,\ell_+]$ in units of $a_0$ is obtained by drawing horizontal
and vertical line segments from it to the diagonal dashed line; its
projection on that line (small black dot) yields
\beq
\ell \approx (\ell_++\ell_-)/2\,,
\eqlab{loc-approx}
\eeq
which we will refer to as the ``strong bound'' estimate, as opposed to
the ``weak bound'' estimate obtained by replacing $\ell_+$ with
$\ell_{++}$ in the expression above.

\begin{figure}
\centering
\includegraphics[width=0.6\columnwidth]{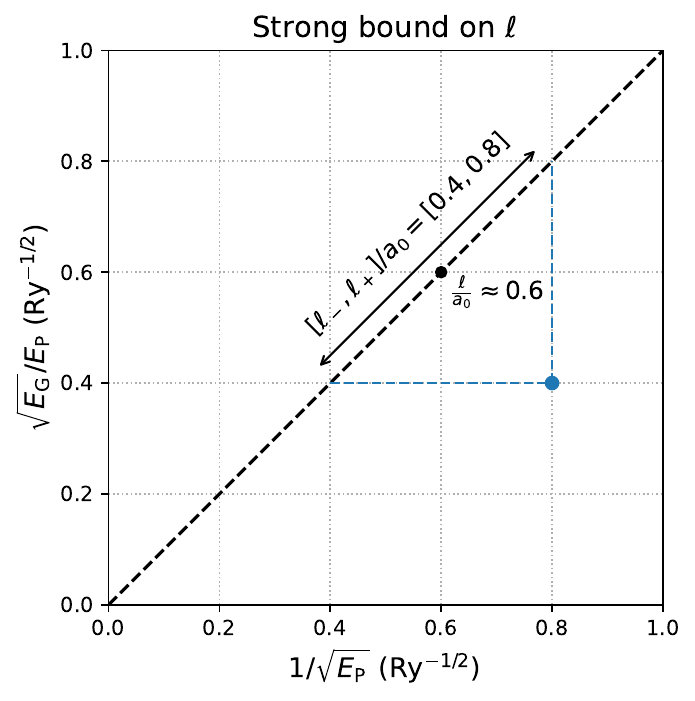}
\caption{Schematic representation of the strong bound on $\ell$ in
  \eq{bounds-loc-fig}.  For the weak bound, replace $\Ep\rightarrow\Eg$ on
  the horizontal axis; since $\Eg\leq\Ep$, the data point (blue dot)
  will move to the right, resulting in a wider range
  $[\ell_-,\ell_{++}]$.}
\figlab{schematic}
\end{figure}

In the following, we use \eqs{bounds-loc-fig}{loc-approx} to estimate
the electron localization length in different classes of materials.
The needed experimental data are the optical gap (the lowest energy
for optical absorption), the electron density, and the clamped-ion
electric susceptibility; the last two enter via \eq{chi} for the Penn
gap.

\subsection{Rocksalt alkali halides}

\begin{figure*}
\centering
\begin{minipage}{0.4\textwidth}
\includegraphics[width=\textwidth]{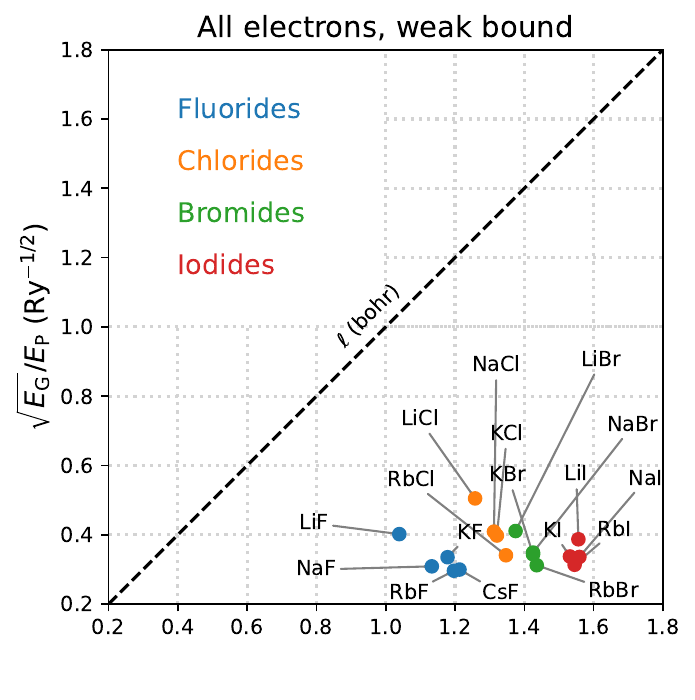}
\end{minipage}
\begin{minipage}{0.4\textwidth}
\includegraphics[width=\textwidth]{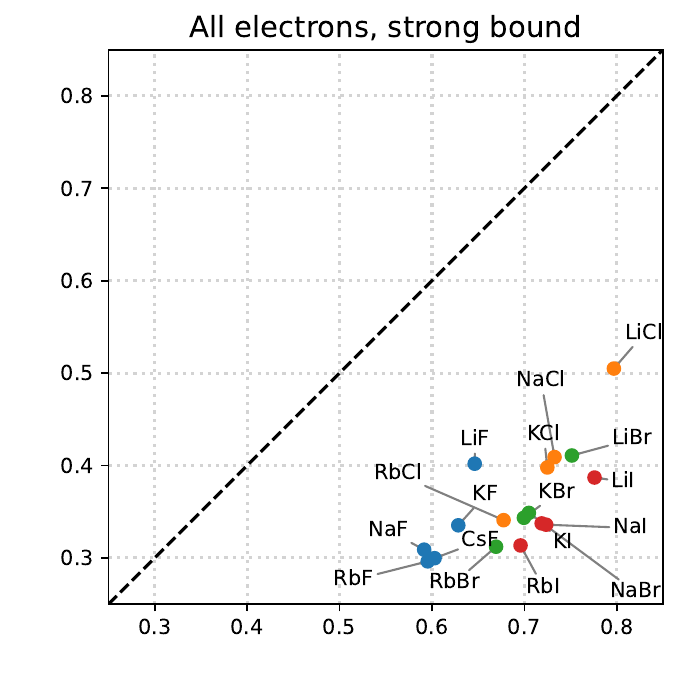}
\end{minipage}
\begin{minipage}{0.4\textwidth}
\includegraphics[width=\textwidth]{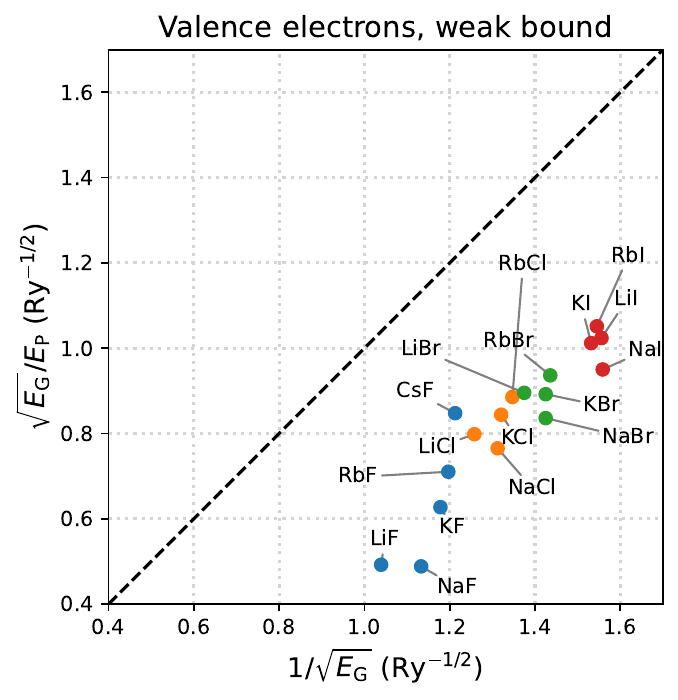}
\end{minipage}
\begin{minipage}{0.4\textwidth}
\includegraphics[width=\textwidth]{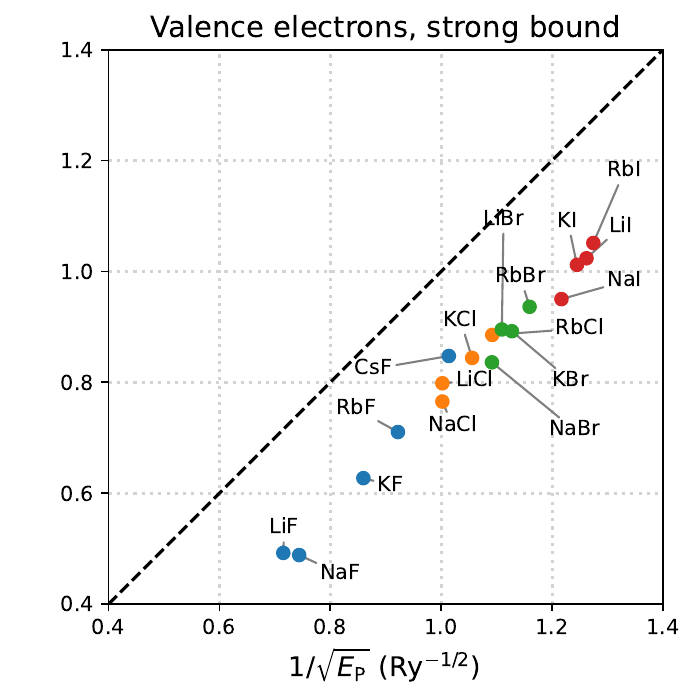}
\end{minipage}
\caption{Bounds on $\ell$ for the rocksalt alkali halides, plotted
  using the scheme outlined in \fref{schematic}.  The weak and strong
  bounds are represented on the left and right panels, respectively,
  while the top and bottom panels show all-electron and
  valence-electron results, respectively, with $\Ep$ defined
  accordingly in each case.}
\figlab{alkali-halides}
\end{figure*}

\Fref{alkali-halides} shows the results obtained for alkali halides
with the rocksalt structure.  Consider first the top panels, where
$\Ep$ was calculated from the total electron density $\nel$ including
inner core electrons.  Such ``all-electron'' bounds inevitably provide
average localization lengths that include those tight inert states; as
a consequence, the bounds are rather loose not only on the left panel
(weak bound) but also on the right panel (strong bound). On the left
panel the upper bound is independent of $\nel$, and hence it is
insensitive to the different localization lengths of valence and core
electrons.  This is not the case for the right panel, where the
tighter upper bound containing $\nel$ narrows down the estimates for
$\ell$; nevertheless, the data points are still quite far from the
diagonal.

To rationalize the results for the strong bound, note that
\beq
\frac{\ell_+ - \ell_-}{a_0} =
\sqrt{\frac{\Ry}{\Ep}}
\left( 1-\sqrt{\frac{\Eg}{\Ep}} \right)\,,
\eqlab{range}
\eeq
and thus the range $[\ell_-,\ell_+]$ gets tighter and tighter as $\Ep$
gets closer to $\Eg$. Since $\Ep\propto\sqrt{\nel}$, the inclusion of
core electrons goes in the opposite direction, and the range
$[\ell_-,\ell_+]$ tends to increase as we move down the periodic
table. This can be seen in the top-right panel of
\fref{alkali-halides}, where the distance from the diagonal line
increases from the fluorides to the chlorides, from these to the
bromides, and from these to the iodides.

It would be much more relevant for physical properties if one could
estimate the average localization length of the valence electrons
only.  Here, we take the simple approach of replacing $\Ep$ in
\eq{bounds-loc-fig} with a valence Penn gap calculated from the
valence electron density.\footnote{Such replacement assumes that
  valence-only versions of the sum rules in \eq{sum-rules} can be
  formulated, which requires the excitation energies of core electrons
  to be well separated in energy from those of valence
  electrons~\cite{wooten-book72}.}  The bounds on $\ell$ obtained in
this manner are presented in the bottom panels of
\fref{alkali-halides}.  As a result of discarding the core electrons
the data points move closer to diagonal line (the bounds get tighter),
and their projections on that line move further up (the average
localization lengths increase). Most interestingly, simple trends
emerge in this valence-only formulation, with $\ell$ increasing from
the lighter to the heavier halogens; this agrees with the intuition on
chemical bonding in strongly ionic
crystals~\cite{ashcroft-mermin-book76}.  The trend is most visible in
the bottom right panel displaying the strong bound. The valence-only
values for $\ell_-$, $\ell_+$, and $\ell_{++}$ are compiled in
Table~\ref{tab:alkali-halides}.

\begin{table}
\centering
\begin{tabular}{cccccccc}
\hline\hline
Crystal & $a\,(\mathring {\mathrm A})$ & $\epsilon$ &
$\Eg$ (eV) & $\ell_-\,(a_0)$ & $\ell_+\,(a_0)$ & $\ell_{++}\,(a_0)$ & \\
\hline
LiF  &  4.02 & 1.96 & 12.6  & 0.40 & 0.65 & 1.04\\
NaF  &  4.62 & 1.74 & 10.6  & 0.31 & 0.59 & 1.13\\
KF   &  5.35 & 1.85 &  9.8  & 0.34 & 0.63 & 1.18\\
RbF  &  5.64 & 1.96 &  9.5  & 0.30 & 0.60 & 1.20\\
CsF  &  6.01 & 2.16 &  9.25 & 0.30 & 0.60 & 1.21\\
\hline
LiCl &  5.13 & 2.78 &  8.6  & 0.50 & 0.80 & 1.26\\
NaCl &  5.64 & 2.34 &  7.9  & 0.41 & 0.73 & 1.31\\
KCl  &  6.29 & 2.19 &  7.8  & 0.40 & 0.72 & 1.32\\
RbCl &  6.58 & 2.19 &  7.5  & 0.34 & 0.68 & 1.35\\
\hline
LiBr &  5.50 & 3.17 &  7.2  & 0.41 & 0.75 & 1.37\\
NaBr &  5.97 & 2.59 &  6.7  & 0.35 & 0.70 & 1.43\\
KBr  &  6.60 & 2.34 &  6.7  & 0.34 & 0.70 & 1.43\\
RbBr &  6.58 & 2.19 &  7.5  & 0.31 & 0.67 & 1.44\\
\hline
LiI  &  6.00 & 3.80 & 5.62  & 0.39 & 0.78 &  1.56\\
NaI  &  6.47 & 2.93 &  5.6  & 0.34 & 0.72 &  1.56\\
KI   &  7.07 & 2.62 &  5.8  & 0.34 & 0.72 &  1.53\\
RbI  &  7.34 & 2.59 &  5.7  & 0.31 & 0.70 & 1.54\\
\hline
\end{tabular}
\caption{Bounds on $\ell$ for the valence electrons in rocksalt alkali
  halides, estimated from experimental data: lattice constant $a$ (the
  electron density is $\nel=32/a^3$, corresponding to eight valence
  electrons per formula unit), electronic permittivity $\epsilon$, and
  optical gap $\Eg$.  The values for $a$ and $\epsilon$ are from
  Ref.~\cite{ashcroft-mermin-book76}, and those for $\Eg$ correspond
  to the lowest absorption peaks in Ref.~\cite{teegarden-pr67}; the
  exceptions are LiF and LiI, for which $\Eg$ are the excitonic gaps
  reported in Refs.~\cite{roessler-jpcs67} and~\cite{bachrach-pla79},
  respectively.}
\label{tab:alkali-halides}
\end{table}

\subsection{Tetrahedrally-coordinated materials}

\begin{figure*}
\centering
\begin{minipage}{0.4\textwidth}
\includegraphics[width=\textwidth]{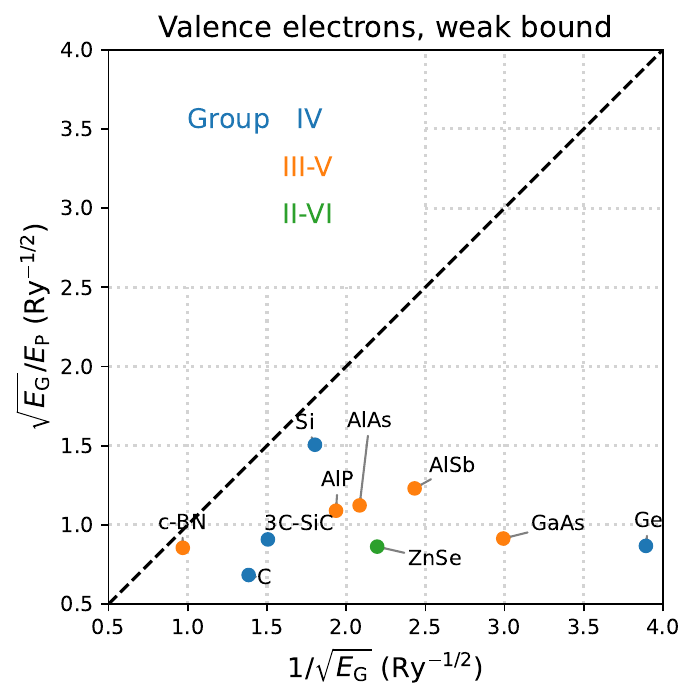}
\end{minipage}
\begin{minipage}{0.4\textwidth}
\includegraphics[width=\textwidth]{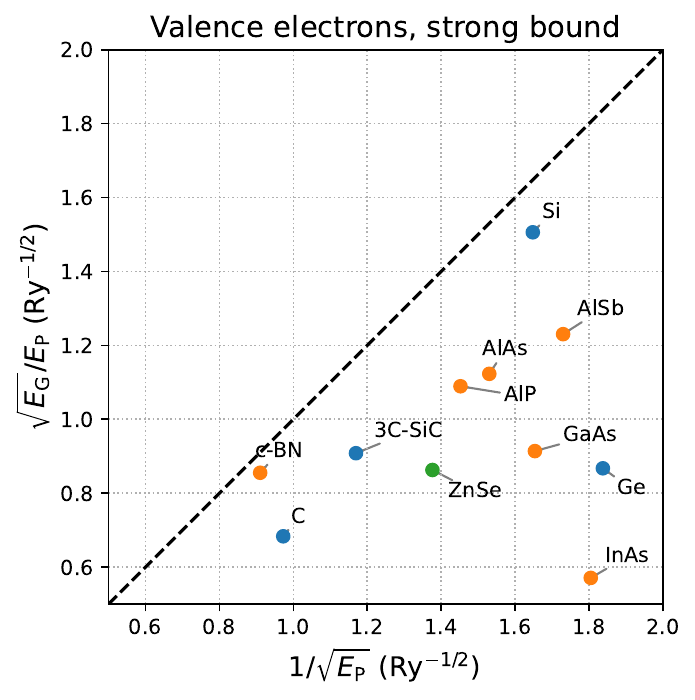}
\end{minipage}
\caption{Same as the bottom two panels of \fref{alkali-halides}, but
  for materials with the diamond or the zincblende structure. On the
  left panel, the data point for InAs is out of bounds.}
\figlab{tetrahedral}
\end{figure*}

\Fref{tetrahedral} and Table~\ref{tab:tetrahedral} show the
valence-only results obtained for materials with the diamond or the
zincblende structure from groups IV, III-V, and II-VI in the periodic
table. The trends are not as uniform as in the case of the halides
because there is a larger range of gaps and
susceptibilities. Nevertheless, one observes that the values of $\ell$
estimated from \eq{loc-approx} tend to decrease with increasing
ionicity, e.g., along the isoelectronic series Si~$\rightarrow$~AlP
and Ge~$\rightarrow$~GaAs~$\rightarrow$~ZnSe, as also found in
Ref.~\cite{sgiarovello-prb01}; this is consistent with the intuition
that ionic bonding yields more localized electrons than covalent
bonding. Accordingly, the estimated localization lengths in
Table~\ref{tab:tetrahedral} tend to be larger than those in
Table~\ref{tab:alkali-halides} for the strongly ionic alkali halides.

\begin{table}
\centering
\begin{tabular}{cccccccc}
\hline\hline
Crystal & $a\,(\mathring {\mathrm A})$ & $\epsilon$ &
$\Eg$ (eV) & $\ell_-\,(a_0)$ & $\ell_+\,(a_0)$ & $\ell_{++}\,(a_0)$ & \\
\hline
C      & 3.57 &  5.7  & 7.1  & 0.68 & 0.97 & 1.38\\
Si     & 5.43 & 11.97 & 4.19 & 1.51 & 1.65 & 1.80\\
Ge     & 5.66 & 16.00 & 0.90 & 0.87 & 1.84 & 3.89\\
3C-SiC & 4.36 &  6.38 & 6.0  & 0.91 & 1.17 & 1.51\\
\hline
c-BN   & 3.62 &  4.46 & 14.5 & 0.86 & 0.91 & 0.97\\
AlP    & 5.46 &  7.5  & 3.63 & 1.09 & 1.45 & 1.94\\
AlAs   & 5.66 &  8.2  & 3.13 & 1.12 & 1.53 & 2.08\\
AlSb   & 6.14 & 10.24 & 2.3  & 1.23 & 1.73 & 2.43\\
GaAs   & 5.65 & 10.86 & 1.52 & 0.91 & 1.65 & 2.99\\
InAs   & 6.06 & 12.37 & 0.42 & 0.57 & 1.80 & 5.71\\
\hline
ZnSe   & 5.68 & 5.7   & 2.82 & 0.86 & 1.38 & 2.20\\
\hline
\end{tabular}
\caption{Same as Table~\ref{tab:alkali-halides}, but for materials
  with the diamond or the zincblende structure. We assume four valence
  electrons per atom on average, so that $\nel=32/a^3$. The
  experimental data is from Ref.~\cite{madelung-book04}, where $\Eg$
  is the direct gap.}
\label{tab:tetrahedral}
\end{table}

How well do the $\ell$ values estimated from experimental data via
\eq{loc-approx} compare with those obtained from first-principles
calculations? To address this question, in \fref{theory-expt} we
compare them with the {\it ab initio} values reported in
Ref.~\cite{sgiarovello-prb01}.  The correlation is quite satisfactory,
although the theoretical values tend to be somewhat larger.  To
explain this trend, one could invoke the band gap underestimation in
density functional theory, which may well lead to a systematic
overestimation of the calculated localization lengths.  Since
expressing $\ell$ as the average of $\ell_-$ and $\ell_+$ is an
approximation, however, it is difficult to draw definitive
conclusions, especially for cases like Ge where $\ell_-$ and $\ell_+$
are rather different.  Yet, it is interesting to note that the upper
bounds in \fref{theory-expt} essentially fall on the diagonal in all
cases, which means that they closely match the available theoretical
data. (The lower bounds, involving the minimum gap, display a much
larger scatter.)  This gives further credit to our earlier statements
that $\ell_+$ is a more robust indicator of the polarization
fluctuation amplitude compared to $\ell_-$.

\begin{figure}
\centering
\includegraphics[width=0.6\columnwidth]{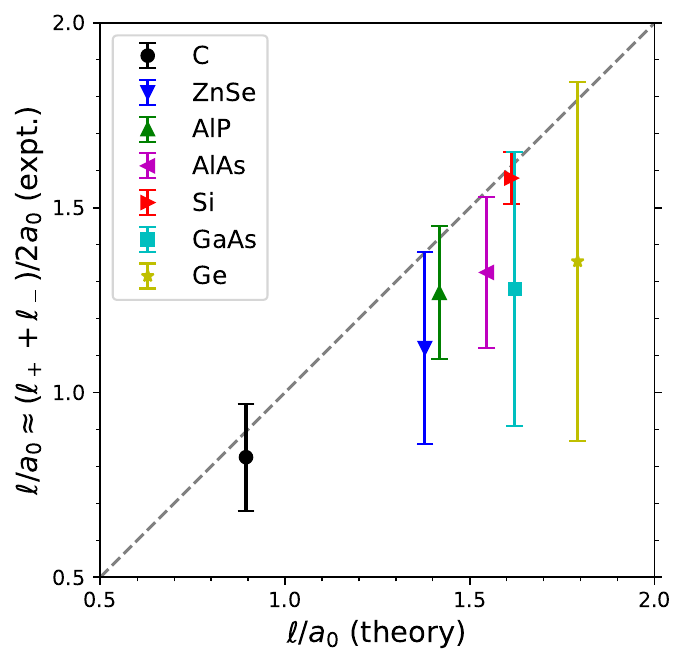}
\caption{Comparison, for tetrehedrally-cordinated materials, between
  the $\ell$ values for valence electrons estimated from experimental
  data (Table~\ref{tab:tetrahedral}), and those calculated from first
  principles in Ref.~\cite{sgiarovello-prb01} using a pseudopotential
  method. The error bars indicate the range
    $[\ell_-,\ell_+]$.}
\figlab{theory-expt}
\end{figure}

\section{Conclusions}
\seclab{conclusions}

The use of sum-rule inequalities to estimate the electronic
polarizability is well established in atomic and molecular
physics~\cite{traini-ejp96,atkins-book05}. The extension of those
ideas to crystals and to other physical properties is not equally
developed, and the results are scattered in the literature. In this
work, we provided a unified perspective on several sum-rule
inequalities for bulk systems, and organized them into chained
inequalities providing bounds on three electronic properties of
insulators: localization length $\ell$, static susceptibility $\chi$,
and optical gap $\Eg$.  As they are based on exact sum rules, those
inequalities remain valid for correlated, disordered, and topological
insulators, and in the presence of relativistic effects including
spin-orbit coupling.  The extension to low-symmetry crystals with
anisotropic localization and susceptibility tensors is also
straightforward.  As an application, we estimated $\ell$ (a
ground-state property) from readily available experimental data on the
response properties $\chi$ and $\Eg$, together with the electron
density. By focusing on the valence electrons, we obtained meaningful
estimates for their average localization length that follow simple
chemical trends.

The study of several exactly solvable models, from the hydrogen atom
to isolated and coupled oscillators, provided useful insights.  In
particular, the coupled oscillator models illustrated how the
fluctuation-dissipation relation breaks down at the mean-field level
and critically requires an explicit treatment of dynamical
correlations.

\section*{Acknowledgements}
We are indebted to Morrel H. Cohen for an unpublished collaboration
with one of us (I.~S.), on which Problem 22.5 of
Ref.~\cite{martin-book04} was based.  We also wish to thank Liang Fu,
Yugo Onishi, and Raffaele Resta for stimulating
discussions. M.~S. thanks the CCQ at the Flatiron Institute for
hospitality while this research was carried out. The Flatiron
Institute is a division of the Simons Foundation.


\paragraph{Funding information}
Work by I.~S. was supported by Grant No.  PID2021-129035NB-I00 funded
by MCIN/AEI/10.13039/501100011033 and by ERDF/EU.  Work by M.~S. was
supported by the State Investigation Agency through the Severo Ochoa
Programme for Centres of Excellence in R\&D (CEX2023-001263-S), and
from Generalitat de Catalunya (Grant No. 2021 SGR 01519).


\begin{appendix}
\numberwithin{equation}{section}

\section{Polarization and localization in band insulators}
\seclab{band-insulators}

In \eqs{Pel-def}{loc-def}, the electronic polarization and the
electron localization tensor were written down for a generic bulk
insulator (possibly correlated and/or disordered) using Kohn's
center-of-mass operator. Alternatively, those expressions can be
recast in terms of the Berry phase and quantum metric defined by the
change in the many-body ground state under twisted boundary
conditons~\cite{ortiz-prb94,souza-prb00}.  Here we specialize to the
single-particle picture, and review the corresponding formulas for
uncorrelated crystalline insulators.

The electronic polarization of a band insulator takes the form of a
Berry phase of the cell-periodic Bloch states in momentum
space~\cite{king-smith-prb93,vanderbilt-book18},
\beq
{\bf P}_{\rm e}=\frac{-|e|}{(2\pi)^3}\int d^3k\,
\sum_{n=1}^J\,{\bf A}_{nn}(\k)\,;
\eqlab{P-berry}
\eeq
here ${\bf A}_{mn}(\k)=i\ip{u\bmk}{{\boldsymbol\nabla}_\k u\bnk}$ is
the Berry connection matrix, the integral is over the first Brillouin
zone (BZ), and the summation is over the valence bands.
Alternatively, ${\bf P}_{\rm e}$ can be written
as~\cite{king-smith-prb93,vanderbilt-book18}
\beq
{\bf P}_{\rm e}=\frac{-|e|}{\Omega}\sum_{n=1}^J\,\ev{\r}_n\,,
\eqlab{P-wannier}
\eeq
where $\ev{\r}_n$ is the center of charge of a Wannier function
constructed for band $n$.

The localization tensor can be obtained from the
fluctuation-dissipation relation in \eq{fluct-diss}. Using the
Kubo-Greenwood formula for the optical conductivity, one finds the sum
rule~\cite{souza-prb00,souza-prb08}
\beq
\int_0^\infty d\w\,
\w^{-1}\,\Re\,\sigma^{\rm S}_{ab}(\w)=
\frac{\pi e^2}{(2\pi)^3\hbar}\int d^3k\,
\sum_{n=1}^J\,g_{ab,nn}(\k)\,.
\eqlab{SWM}
\eeq
On the right-hand side, $g(\k)$ is the quantum metric
tensor~\cite{provost80} of the valence
manifold~\cite{marzari-prb97,vanderbilt-book18},
\beq
g_{ab,mn}(\k)=
\frac{1}{2}\me{\partial_a u\bmk}{Q_\k}{\partial_b u\bnk}+
\frac{1}{2}\me{\partial_b u\bmk}{Q_\k}{\partial_a u\bnk}\,,
\eeq
with $\partial_a=\partial/\partial k_a$ and
$Q_\k=\mathbbm{1}-\sum_{n=1}^J\,\ket{u\bnk}\bra{u\bnk}$.  Inserting
\eq{SWM} in \eq{fluct-diss} gives
\beq
\ell^2_{ab}=
\frac{1}{(2\pi)^3\nel}
\int d^3k\,\sum_{n=1}^J\,g_{ab,nn}(\k)\,,
\eqlab{loc-metric}
\eeq
which expresses the bulk localization tensor as a ground-state
quantity.

For a one-dimensional (1D) insulator the localization tensor reduces
to a scalar, and \eq{loc-metric} can be written in terms of
maximally-localized Wannier functions as
\beq
\ell^2=\frac{1}{J}\sum_{n=1}^J\,
\left(
\ev{x^2}_n-\ev{x}_n^2
\right)\,,
\eqlab{loc-WF}
\eeq
which follows from the relation between the BZ integral of the metric
and the quadratic Wannier spread~\cite{marzari-prb97}.  Thus, in 1D
the localization tensor is equal to the average spread of the
maximally-localized Wannier functions.  More generally, in $d$
dimensions its Cartesian trace equals the gauge-invariant part of the
average Wannier spread, which for $d>1$ is smaller than the actual
spread in any gauge~\cite{marzari-prb97}.

In summary, electronic polarization is related to the Wannier centers
of the valence bands, and the electron localization length squared
(polarization fluctuations) gives a lower bound to the average Wannier
spread.

\section{Longitudinal optical bounds}
\seclab{longitudinal}

Here, we outline the extension to long-wave longitudinal
modes~\cite{onishi-prb24,onishi2024quantum} of the analysis carried
out in \srefs{sum-rules}{bounds} for transverse modes.  We again
assume cubic symmetry or higher so that
$\epsilon_{ab}(\w)=\delta_{ab}\epsilon(\w)$, and define the moments of
the energy-loss spectrum as
\beq
M_p=\frac{2}{\pi}\int_0^\infty d\w\, \w^p\,
\Im \left[ -\epsilon^{-1}(\w) \right]\,.
\eqlab{M-p}
\eeq
The moments with $p=1,0,-1$ satisfy the relations
\begin{subequations}
\begin{align}
M_1 &= \wpl^2\,,\eqlab{M1-SR}\\
M_0 &= \frac{2e^2}{\hbar\epsilon_0}\nel\tilde\ell^2\,,\eqlab{M0-SR}\\
M_{-1} &= 1-\epsilon^{-1}\,,\eqlab{M-1-SR}
\end{align}
\end{subequations}
where $\epsilon^{-1}$ stands for $\epsilon^{-1}(0)$.  These are
respectively the longitudinal counterpart of the oscillator-strength
sum rule~\eqref{eq:I0-SR}~\cite{wooten-book72}, the longitudinal
fluctuation-dissipation relation~\eqref{eq:fluct-diss-longitudinal},
and the longitudinal counterpart of the Kramers-Kr\"onig
relation~\eqref{eq:I2-SR}.

Next, we introduce average gaps for longitudinal excitations by
analogy with \eqr{El-Ep}{chi},
\beq
\tilde E_{\rm L}=\frac{\hbar M_1}{M_0}\,,
\qquad
\tilde E_{\rm P}^2=\frac{\hbar M_1}{\hbar^{-1} M_{-1}}\,,
\eqlab{El-Ep-longitud}
\eeq
\beq
\tilde E_{\rm L}=\frac{\hbar^2}{2\mel\tilde\ell^2}
\Leftrightarrow
\left( \frac{\tilde\ell}{a_0} \right)^2 = \frac{\Ry}{\tilde E_{\rm L}}
\eqlab{loc-longitud}
\eeq
\beq
1-\epsilon^{-1}=\left( \frac{\hbar\wpl}{\tilde E_{\rm P}}\right)^2\,.
\eqlab{chi-longitud}
\eeq
Since the loss function appearing in \eq{M-p} is positive
semidefinite, one can immediately write down inequalities analogous to
those in Eqs.~\eqref{eq:Eg-ineqs}, \eqref{eq:cauchy-ineq},
and~\eqref{eq:El-Ep-ineqs},
\beq
M_{p-q}\leq\frac{\hbar}{\tilde E_{\rm G}}M_{p-q+1}\leq\ldots\leq
\left( \frac{\hbar}{\tilde E_{\rm G}} \right)^q M_p
\eqlab{Eg-ineqs-longitud}
\eeq
($\tilde E_{\rm G}$ is the plasmon gap),
\beq
M_p^2\leq M_{p-1}M_{p+1}\,,
\eqlab{cauchy-ineq-longitud}
\eeq
and
\beq
\tilde E_{\rm L} \geq \tilde E_{\rm p} \geq \tilde E_{\rm G}\,,\qquad
\tilde E_{\rm P}^2 \geq \tilde E_{\rm G} \tilde E_{\rm L}\,.
\eqlab{El-Ep-ineqs-longitud}
\eeq

Finally, by forming the chained inequalities
\begin{subequations}
\begin{align}
\tilde E_{\rm P}^2/\tilde E_{\rm G} &\geq \tilde E_{\rm L} \geq
\tilde E_{\rm P} \geq \tilde E_{\rm G}\,,\eqlab{ineqs-El-longitud}\\
\tilde E_{\rm L}^2 &\geq \tilde E_{\rm P}^2 \geq
\tilde E_{\rm G} \tilde E_{\rm L} \geq \tilde E_{\rm G}^2\,,\\
\tilde E_{\rm G} &\leq \tilde E_{\rm P}^2/\tilde E_{\rm L} \leq
\tilde E_{\rm P} \leq \tilde E_{\rm L}
\end{align}
\end{subequations}
and combining them with \eqs{loc-longitud}{chi-longitud}, we obtain
weak and strong bounds on $\tilde\ell^2$, $1-\epsilon^{-1}$ and
$\tilde E_{\rm G}$, respectively. Those bounds are given by
\eqs{bounds-loc-chi}{bounds-gap}, with the replacements indicated in
\eq{changes-LO}.

\section{Zero-point energy of the van der Waals crystal model}
\seclab{vdw-zpe}

In this appendix we return to the van der Waals crystal model of
\sref{vdw-xtal}, and calculate its zero-point energy in two different
ways. First we use a Brillouin-zone integral,
\begin{equation}
E = \frac{\hbar}{2}
\frac{\Omega}{(2\pi)^3} \int d^3 q \sum_i\, \w_i({\bf q})\,.
\eqlab{zpe-BZ}
\end{equation}
To verify that the normalization factors are correct, note that in the
absence of interactions we recover the correct result for the isolated
3D oscillator,
\begin{equation}
E_0 = \frac{3}{2}\hbar \w_0\,.
\end{equation}
The interaction is regarded as a small perturbation, so we can 
Taylor-expand the square root of \eq{omega-q} for $\w^2_i(\q)$,
\begin{equation}
\w_i = \sqrt{\w_0^2 + \lambda_i} \simeq \w_0 +
\frac{1}{2}\frac{\lambda_i}{\w_0} - \frac{1}{8}\frac{\lambda^2_i}{\w^3_0}\,.
\eqlab{w-expand}
\end{equation}
As the $C(\q)$ matrix defined by \eq{C-q} is traceless for all $\q$,
the second term above drops out from \eq{zpe-BZ}. The leading
correction is then given by the third term,
\begin{equation}
\Delta E = -\frac{\hbar}{16 \w_0^3} \frac{\Omega}{(2\pi)^3} \int d^3 q
\sum_i\, \lambda^2_i({\bf q})\,.
\eqlab{Delta-E-recip}
\end{equation}
Overall, the interaction energy is negative and in view of
\eq{lambda-TO-LO} it appears to scale as $\Omega^{-2}$, which at first
sight seems consistent with van der Waals. This is confirmed by a
numerical evaluation of \eq{Delta-E-recip} for a simple-cubic lattice
(\fref{converg}), which shows a $\Omega^{-2}$ behavior for $\Delta E$
in the limit of a dense $\q$ mesh.

\begin{figure}
\centering
\includegraphics[width=0.6\columnwidth]{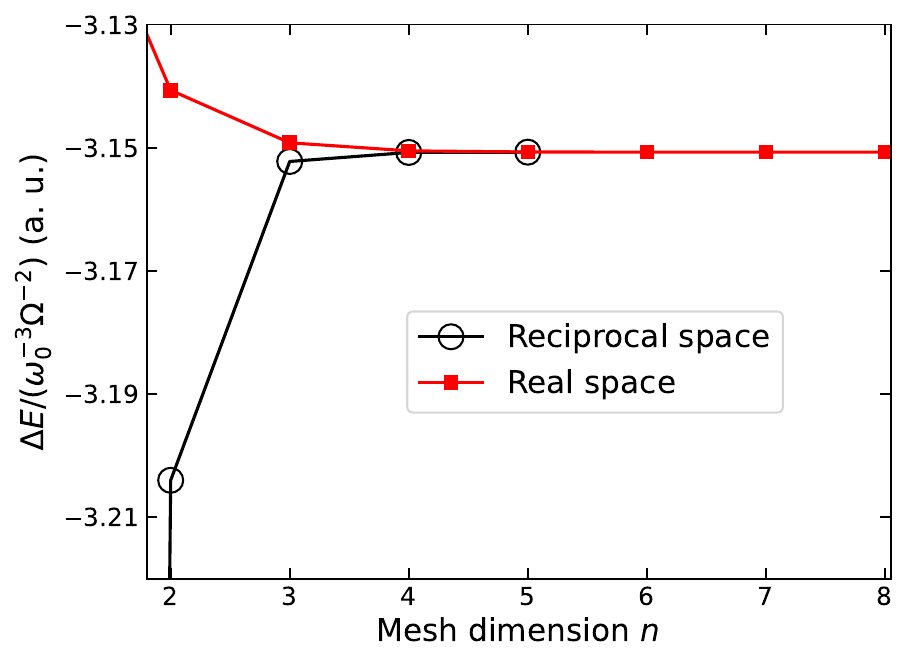}
\caption{Convergence of the reciprocal- and real-space sums for the
  dispersion interaction energy using meshes of dimension $2n$. The
  plotted values correspond to \eqs{Delta-E-recip}{Delta-E-real} for a
  simple-cubic lattice, in units of $\omega_0^{-3}\Omega^{-2}$ using
  Hartree atomic units (a.~u.).}
\figlab{converg}
\end{figure}

As further validation, we have computed the same energy as a
real-space sum of pair interactions. We start from the interaction
energy of the 3D dimer model of \sref{vdw-dimer}, which is obtained by
expanding \eqs{w-parallel}{w-perp} according to \eq{w-expand}.  The
result~\cite{atkins-book05}
\beq
\Delta E_{12} =
-\frac{3}{4}\left( \frac{\alpha'_0}{R^3} \right)^2\hbar\w_0\,,
\eeq
which is enhanced by a factor of $3/2$ relative to that of the 1D
dimer model~\cite{kittel-book04}, leads to a crystal energy of
\beq
\Delta E = -\hbar
\left(
\frac{e^2/\mel}{4\pi\epsilon_0}
\right)^2
\frac{3}{8\w_0^3}\sum_{\R\not=\0}\,\frac{1}{R^6}\,.
\eqlab{Delta-E-real}
\eeq
(Note the additional factor of $1/2$ to avoid double counting of the
pair interactions.)  As shown in \fref{converg}, the converged value
of this real-space summation agrees with that of the reciprocal-space
summation~\eqref{eq:Delta-E-recip}. The plotted quantity is
$\Delta E/(\w_0^{-3}\Omega^{-2})$ in Hartee atomic units, and its
converged value is precisely $-(3/8)A_6$, where
\beq
A_6\equiv \sum_{i,j,k}'\,(i^2+j^2+k^2)^{-3}
\simeq 8.40192
\eeq
(with $i=j=k=0$ excluded) is a lattice sum tabulated by Lennard-Jones
and Ingham~\cite{lennard-jones25}.

\end{appendix}







\end{document}